\documentclass{PoS}

\usepackage{JackCommands}
\usepackage{subcaption}
\usepackage{float}
\usepackage{amsmath}

\title{Improvements to Nucleon Matrix Elements within a $\theta$ Vacuum from Lattice QCD}

\ShortTitle{Improved Matrix Elements in a $\theta$ Vacuum}

\author{\speaker{Jack Dragos}, Andrea Shindler, Ahmed Yousif\\
        Facility for Rare Isotope Beams, Physics Department, Michigan State University, East Lansing, Michigan, USA\\
        E-mail: \email{dragos@frib.msu.edu}}

\author{Thomas Luu\\
       Institute for Advanced Simulation (IAS-4) FZJ, Germany\\}

\author{Jordy de Vries\\
        Amherst Center for Fundamental Interactions, Department of Physics,
        University of Massachusetts Amherst, Amherst, MA, USA \\
        RIKEN BNL Research Center, Brookhaven National Laboratory, Upton, New York, USA\\}

\abstract{
Using the gradient flow, we calculated the nucleon mixing angle $\alphan$ and the nucleon electric dipole moment (EDM)
induced by the QCD $\theta$-term. To do so, we computed the topological charge, and the nucleon two-point and three-point correlation functions.
The purpose of these proceedings is to explore how the topological charge density interacts with the nucleon interpolation operators.
By understanding this relation, we can try to suppress noise contributions to the $\alphan$ and EDM signals by selecting
specific regions where the signal dominates.

Using gauge fields provided by PACS-CS at $N_{f}=2+1$, a first collection of ensembles were selected at a fixed lattice spacing
$a=0.0907$~fm ($\beta=1.90$), fixed dimensions $32^{3}\times 64$ and varying $m_{\pi}\approx\lbrace 411,\,570,\,701\rbrace$~MeV.
A second collection was selected at fixed $m_{\pi}\approx701$~MeV, fixed box size of $L\approx1.9$~fm and
varying $a=\lbrace 0.1215,\,0.0980,\,0.0685\rbrace $~fm.
}

\FullConference{The 36th Annual International Symposium on Lattice Field Theory - LATTICE2018\\
		22-28 July, 2018\\
		Michigan State University, East Lansing, Michigan, USA.}
\begin{document}

\section{Introduction}
The computation of the nucleon EDM from lattice QCD has become quite a priority, as its crucial to understand the ongoing and future neutron and proton EDM experiments.
Although the dimension-four "{\it$\theta$-term}" that is permitted in the Standard Model can induce an EDM,
higher-dimensional operators arising from beyond the Standard Model (BSM) physics could also be responsible.
The computation of the EDM induced specifically by the $\theta$-term helps to disentangle different CP-violating sources from future experimental results.

While the theta term and the link to the nucleon EDMs has been explored in lattice QCD in the past
 \cite{Shintani:2015vsx,Alexandrou:2015spa,Guo:2015tla,Aoki:2008gv,Berruto:2005hg,Shintani:2005xg,Shindler:2015aqa},
 the result have not been satisfactory mainly due to signal-to-noise problems.
 In this spirit, this proceedings explores the relation between the topological
 charge density and
the nucleon interpolating operators in order to improve the signal to noise. This technique
will be employed to the modified nucleon two-point correlation function to improve the determination of the
nucleon mixing angle $\alphan$, as well as the nucleon three-point correlation function to improve the determination of
the proton and neutron EDM.

This signal-to-noise improvement, when used with the gradient flow to overcome divergences and renormalization complications,
makes the $\theta$-term induced continuum extrapolated nucleon EDM from lattice QCD computationally feasible.

\section{$\theta$-term using the Gradient Flow}

The QCD Lagrangian, including the CP-violating $\theta$-term, has the form
\begin{equation}
  \mathcal{L}_{QCD}(x) = \frac{1}{4}\Gmunu^{(a)}(x)G^{(a)\mu\nu}(x) + \sum_{q} \overline{\psi}_{q}(x) \left[\ghmu D_{\mu}(x) - m_{q}\right]\psi_{q}(x)-i\theta q(x),
\end{equation}
where the topological charge density is defined as:
\begin{equation}
  q(x) \equiv
  \frac{1}{32\pi^2}\epsilon_{\mu\nu\rho\sigma}\Trace\left[G^{\mu\nu}(x)G^{\rho\sigma}(x)\right].
\end{equation}

We define the topological charge density using the gradient flow (GF) \cite{Luscher:2010iy}.
The flow time radius \(\sqrt{8t_{f}}\) signifies
how large the radius of smearing is as a result of applying the gradient flow. This constitutes a relabeling of
\begin{equation}
  q(x) \xrightarrow{GF} q(x,t_{f}) = \frac{1}{32\pi^2}\epsilon_{\mu\nu\rho\sigma}\Trace\left[G^{\mu\nu}(x,t_{f})G^{\rho\sigma}(x,t_{f})\right]\,.
\end{equation}
\section{Lattice Parameters}
This study was performed on the PACS-CS gauge fields
available through the ILDG \cite{Beckett:2009cb}. They have \(N_{f}=2+1\), and are
generated using a non-perturbative \(O(a)\)-improved Wilson fermion action
along with an Iwasaki gauge action.
The main ensembles used for this study are of \(32^{3}\times 64\) dimensions
with \(a \simeq 0.0907\) fm (\(\beta = 1.90\)).
The 3 ensembles used have \(m_{\pi} = \{411,\,570,\,701\}\)~MeV, which helps to understand the chiral behavior
of the improvement techniques.

Further studies were performed on lattices of dimensions
\(16^{3}\times32\), \(20^{3}\times40\) and \(28^{3}\times56\),
with lattice spacings of \(a=\{0.1215,0.0980,0.0685\}\) fm. This showed us how effective the improvement
technique is at different lattice resolutions.
Information about all the studied ensembles are contained in Refs.~\cite{Aoki:2008sm,Ishikawa:2007nn}.

\section{Nucleon Mixing Angle \(\alphan(\theta)\)\label{sec:alpha}}
To compute the nucleon mixing angle, the starting point is to understand how the two point correlation
function changes as we move from a CP-even theory to a theory including the $\theta$ term.

The small-$\theta$ expansion applied to the nucleon two-point correlation function
helps us relate expectation values in both theories
\begin{eqnarray}
  G_{2}(\vec{p}^{\, \prime},t,t_f)_{\theta} =
  G_{2}(\vec{p}^{\, \prime},t,\Gamma_+) +
  i\theta G^{(Q)}_{2}(\vec{p}^{\, \prime},t,\Gamma_+\gamma_5,t_{f}) +
  \mathcal{O}(\theta^{2}),
\end{eqnarray}
where the definition of the standard two-point correlation function has the form
\begin{eqnarray}\label{eq:C2}
G_{2}(\vec{p}^{\, \prime},t,\Gamma_+) =
\mathrm{Tr}\left\lbrace\Gamma_+ G_{2}(\vec{p}^{\, \prime},t)\right\rbrace =
\sum_{\vec{x}}
e^{-i\vec{p}^{\, \prime}\cdot\vec{x}}
\,\mathrm{Tr}\left\lbrace \Gamma_+
\braket{ \chi(\vec{x},t)
\overline{\chi}(\vec{0},0)}
\right\rbrace ,
\end{eqnarray}
and we defined a modified two-point correlation function
\begin{eqnarray}\label{eq:C2Q}
G^{(Q)}_{2}(\vec{p}^{\, \prime},t,\Gamma_+\gamma_5,t_{f}) =
\mathrm{Tr}\left\lbrace\Gamma_+\gamma_5 G^{(Q)}_{2}(\vec{p}^{\, \prime},t,t_{f})\right\rbrace =
\sum_{\vec{x}}
e^{-i\vec{p}^{\, \prime}\cdot\vec{x}}
\,\mathrm{Tr}\left\lbrace \Gamma_+\gamma_5
\braket{\chi(\vec{x},t)
\bar{\chi}(\vec{0},0)
Q(t_{f})}
\right\rbrace\,,
\end{eqnarray}
where the topological charge $Q(t_f)$ is the usual space-time integral of the charge density.
In the previous definitions, \(\overline{\chi}\) and \(\chi\) are interpolating operators
with the quantum numbers of a nucleon,
inserted with a source-sink time separation of $t$ and \(\Gamma_{+} = \frac{(I+\gamma_{4})}{2}\).

By working out the spectral decomposition of all three two-point correlators, one finds that the
nucleon mixing angle, up to first order in \(\theta\), $\alphan(\theta) = \alphan^{(1)}\theta +O(\theta^{3})$, can be extracted from the ratio
\begin{equation}\label{eq:alpha}
  \alphan^{(1)} \underset{t \rightarrow \infty}{\sim}
  \frac{
   G^{(Q)}_{2}(0,t,\Gamma_{+}\gamma_{5},t_{f})
   }{
  G_{2}(0,t,\Gamma_{+})
  }\,.
\end{equation}
\vspace*{-5mm}
\begin{figure}[H]
\centering
\begin{subfigure}{.45\textwidth}
  \centering
  \includegraphics[trim={11mm 11mm 12mm 11mm},clip,width=\linewidth]{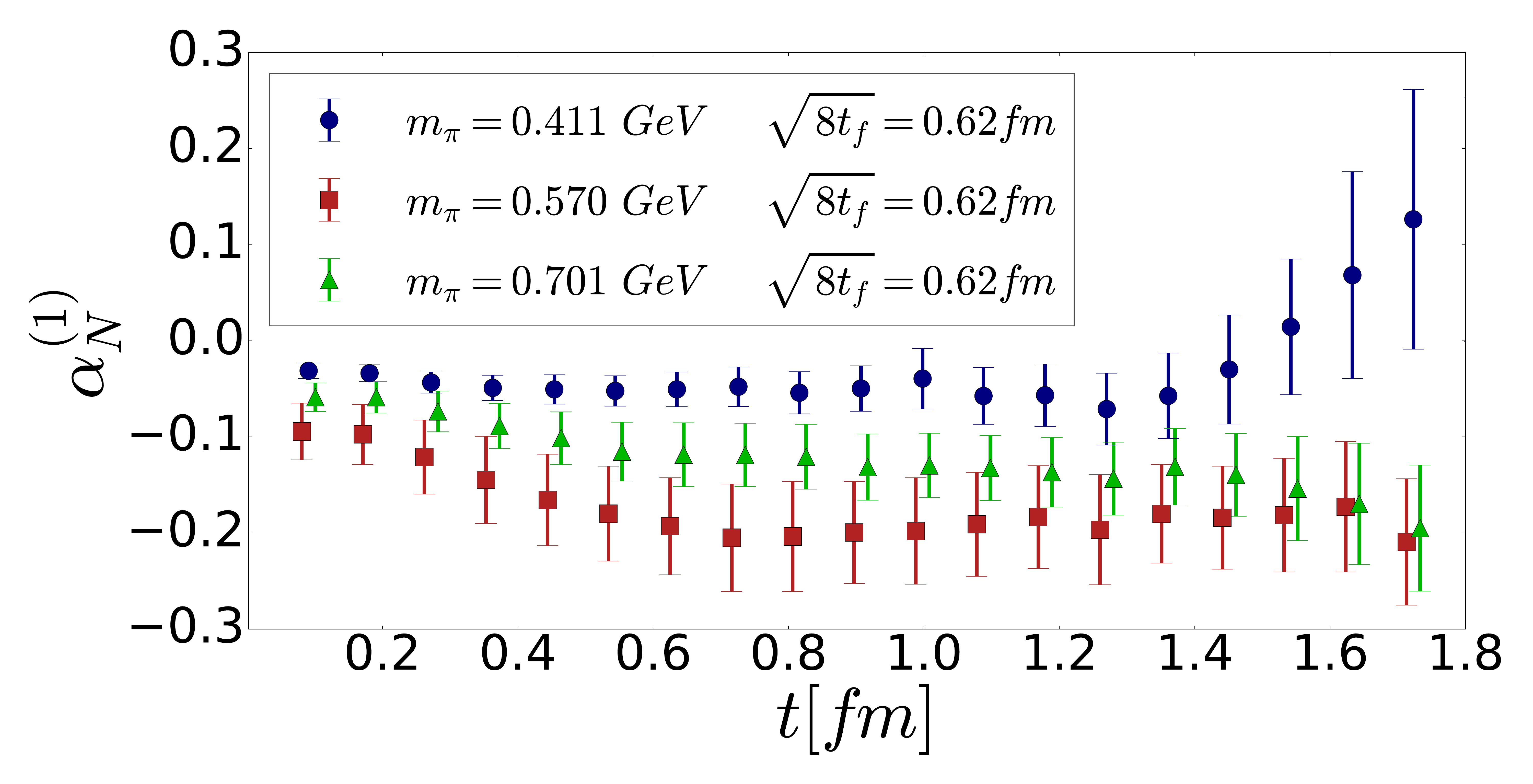}
\end{subfigure}
\quad
\begin{subfigure}{.45\textwidth}
  \centering
  \includegraphics[trim={11mm 11mm 12mm 11mm},clip,width=\linewidth]{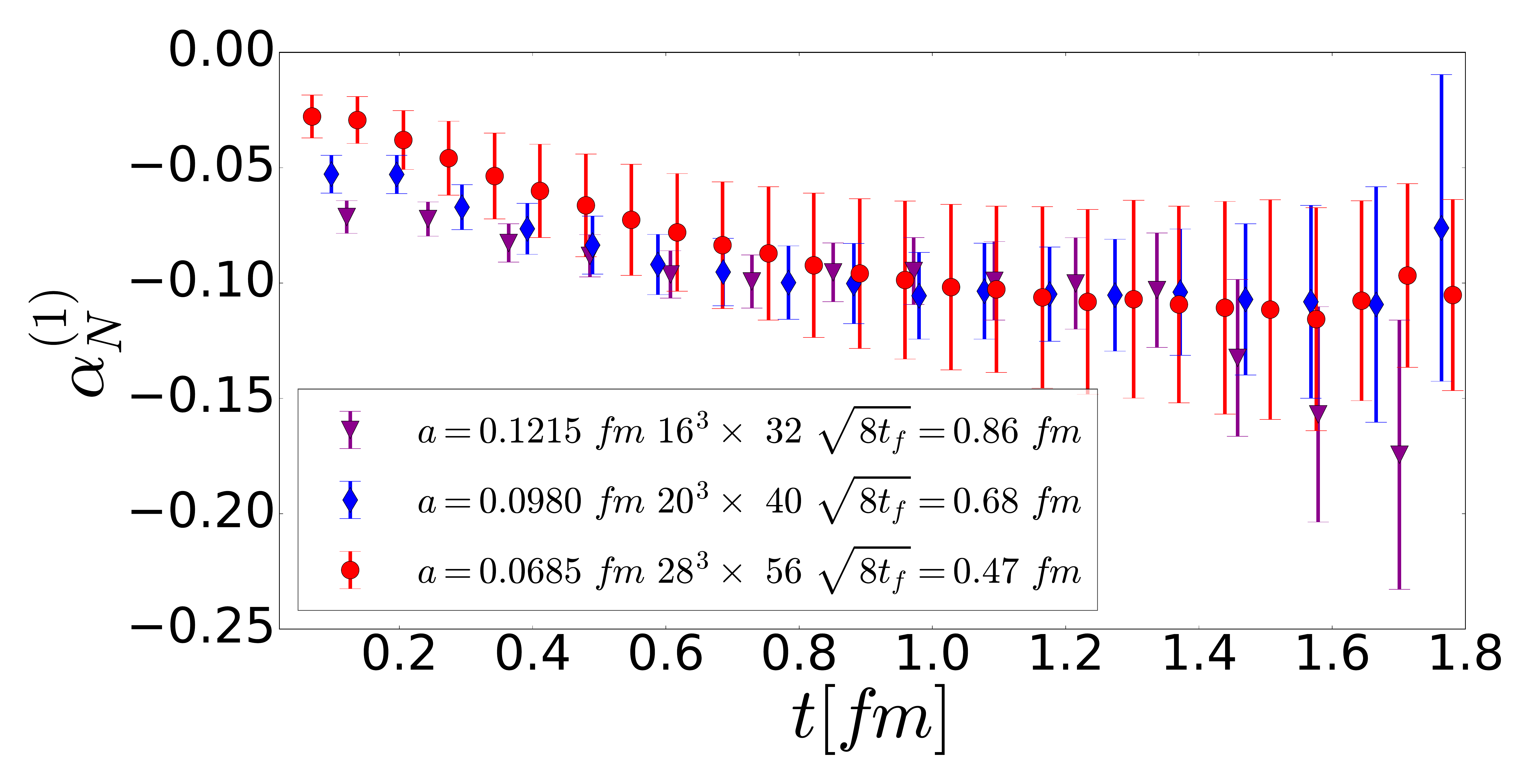}
\end{subfigure}
\caption{
  The $m_{\pi}$ (left) and lattice spacing (right) ensemble results for the nucleon mixing angle \(\alphan^{(1)}\),
  as defined in eq.\ref{eq:alpha}, plotted against source-sink separation \(t\).
  \label{fig:aQ}}
\end{figure}
The standard method for computing the nucleon mixing angle \(\alphan^{(1)}\) is shown in fig.~\ref{fig:aQ}, plotted against
source sink separation \(t\). A plateau to extract $\alphan^{(1)}$ is easily found.
For the ensembles at different lattice spacings in the right plot of fig.~\ref{fig:aQ},
we observe no discretization effects, as the results are statistically consistent
with one another in the range of \(t\) for which a plateau has formed.

\subsection{Improving the Nucleon Mixing Angle \(\alphan(\theta)\)\label{sec:alpha_imp}}

Motivated by a similar study to the topological susceptibility presented in \cite{Aoki:2017paw},
we attempt to improve the signal-to-noise ratio by summing the topological charge density
only over the spatial directions and study the corresponding Euclidean time dependence $\tau$,
In particular we study the dependence of the signal to noise with respect of the time distance
between the charge insertion and the nucleon interpolating operators
\begin{eqnarray}\label{eq:C2QImprov}
G^{(Q)}_{2}(\tau,t,\Gamma_+\gamma_5,t_{f}) =
\sum_{\vec{x}}
\mathrm{Tr}\left\lbrace \Gamma_+\gamma_5\braket{
\chi(\vec{x},t)
Q(\tau,t_{f})
\bar{\chi}(\vec{0},0)
}\right\rbrace \, , \,
Q(\tau,t_{f}) = a^3 \sum_{\vec{y}} q(\vec{y},\tau,t_f)\,.
\end{eqnarray}
Here and in the following the correspondent representations of the correlation functions
as in eq.~\eqref{eq:C2QImprov} in terms of operators expectation values has to be considered
time-ordered.
\begin{figure}[H]
\centering
\begin{subfigure}{.25\textwidth}
  \centering
  \includegraphics[trim={11mm 11mm 11mm 11mm},clip,width=\linewidth]{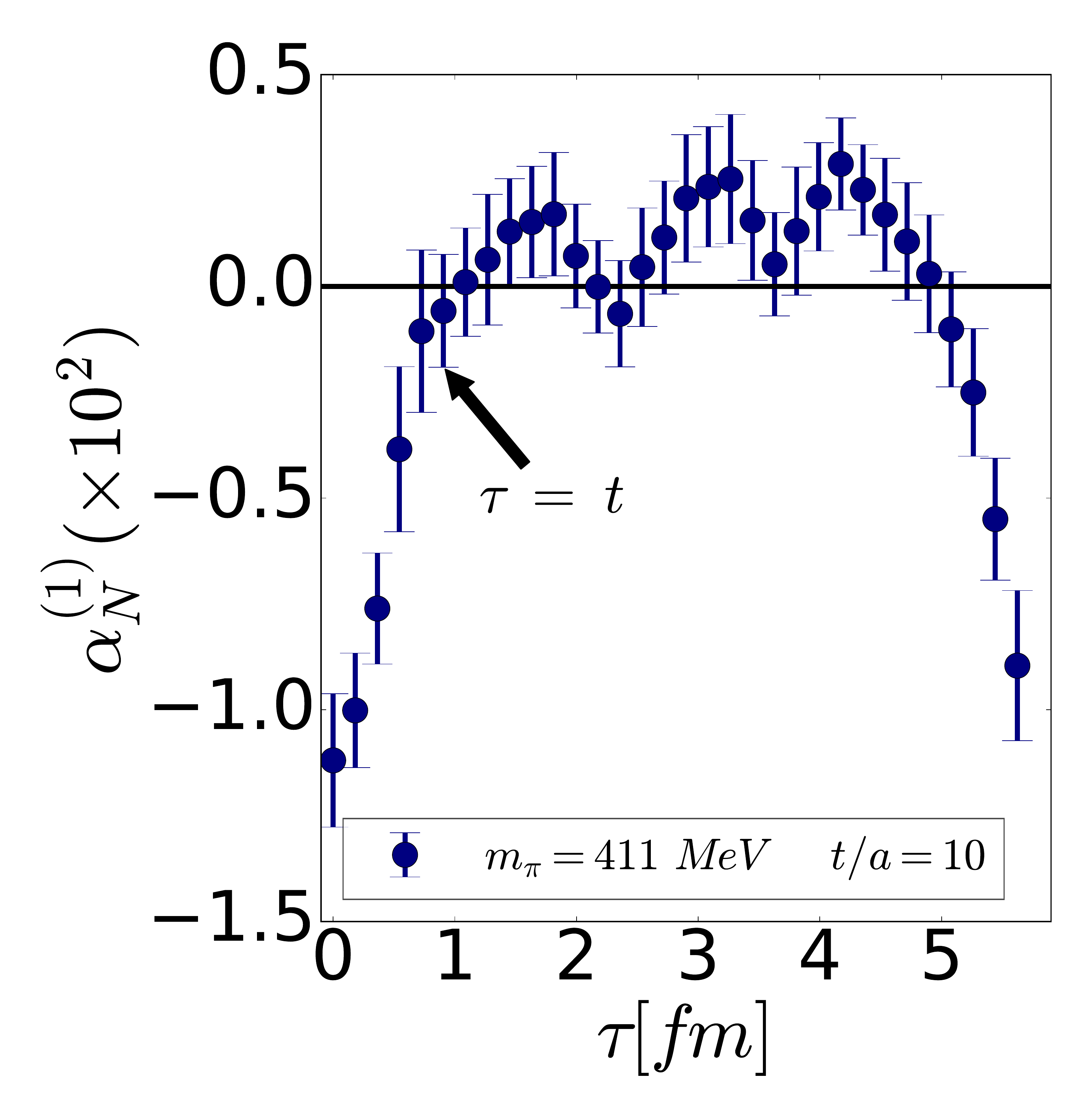}
\end{subfigure}
\qquad
\begin{subfigure}{.25\textwidth}
  \centering
  \includegraphics[trim={11mm 11mm 11mm 11mm},clip,width=\linewidth]{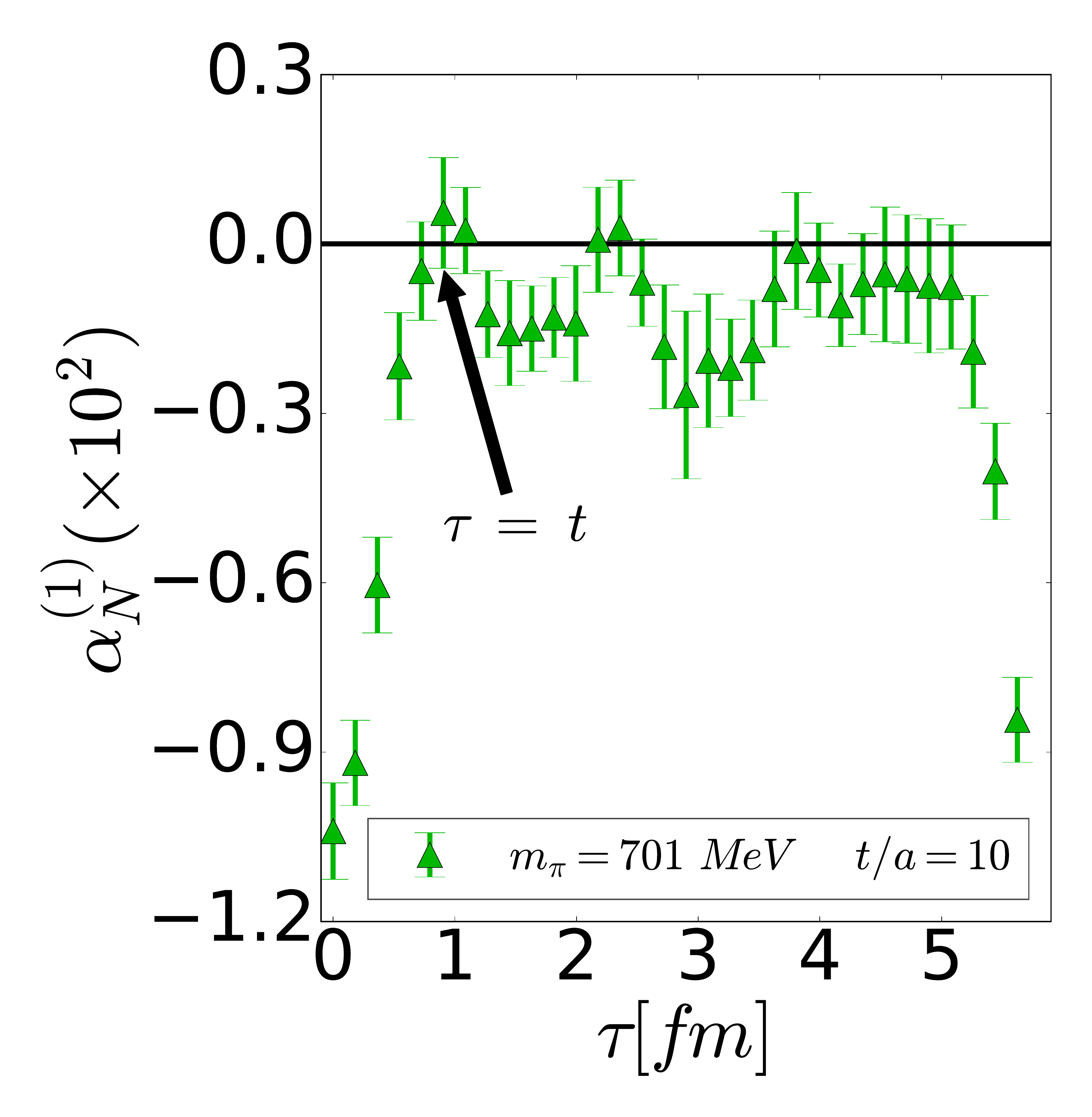}
\end{subfigure}
\qquad
\begin{subfigure}{.25\textwidth}
  \centering
  \includegraphics[trim={11mm 11mm 11mm 11mm},clip,width=\linewidth]{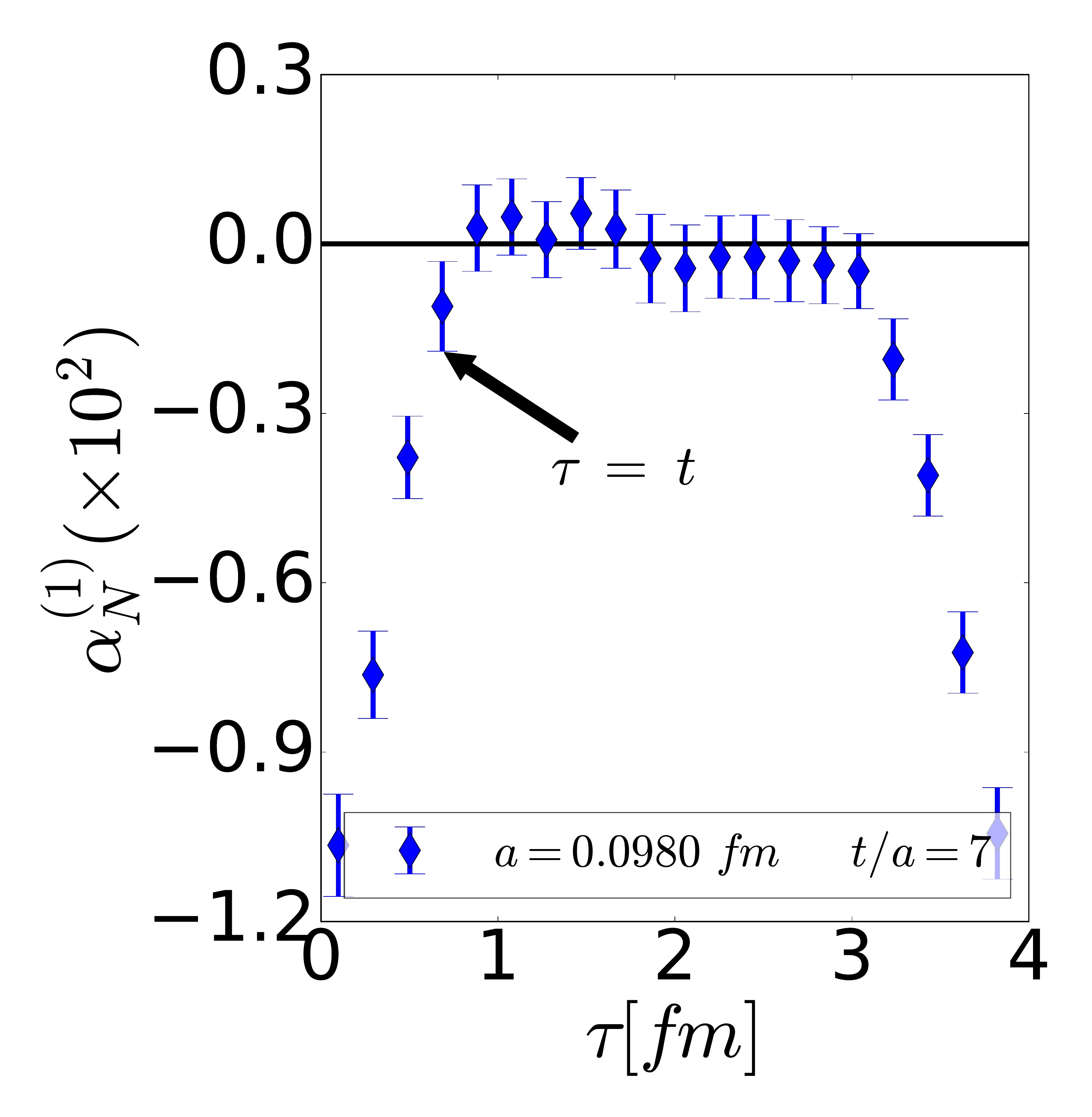}
\end{subfigure}
\caption{\label{fig:a_tcurr}
  The nucleon mixing angle \(\alphan\) computed with a topological charge insertion time \(\tau\) (in fm).
  Every even \(\tau\) is excluded to improve visualization.
  Left, middle and right plots were computed on the \(m_{\pi} = \{411,701\}\)~MeV and
  \(a=0.0980\) fm ensembles.
  The arrow indicates the location of the nucleon interpolating operator at the sink, $t$.
  }
\end{figure}
With the choice of the spin projector, fig.~\ref{fig:a_tcurr} shows how the signal is dominated by the
contribution at \(\tau \rightarrow 0\) 
\footnote{In the following we omit the dependence on the projector of the correlation functions.}.
With this knowledge, one can symmetrically sum about $\tau=0$ and
study the convergence to the total sum
\begin{eqnarray}\label{eq:C2QImprov_2}
&&G^{(Q)}_{2}(t,t_{f},t_{s}) =
\sum_{\tau = 0}^{t_{s}} \left[G^{(Q)}_{2}(\tau,t,t_{f})
+ G^{(Q)}_{2}(T-\tau,t,t_{f})\right] = \nonumber\\ &&
\sum_{\tau = 0}^{t_{s}} \sum_{\vec{x}}
\mathrm{Tr}\left\lbrace \Gamma_{+}\gamma_{5}
\braket{\chi(\vec{x},t)
Q(\tau,t_{f})
\bar{\chi}(\vec{0},0)} +
\Gamma_+\gamma_5
\braket{\chi(\vec{x},t)
Q(T-\tau,t_{f})
\bar{\chi}(\vec{0},0)}
\right\rbrace\,.
\end{eqnarray}
\begin{figure}[H]
\centering
\begin{subfigure}{.45\textwidth}
  \centering
  \includegraphics[trim={11mm 0cm 12mm 0cm},clip,width=\linewidth]{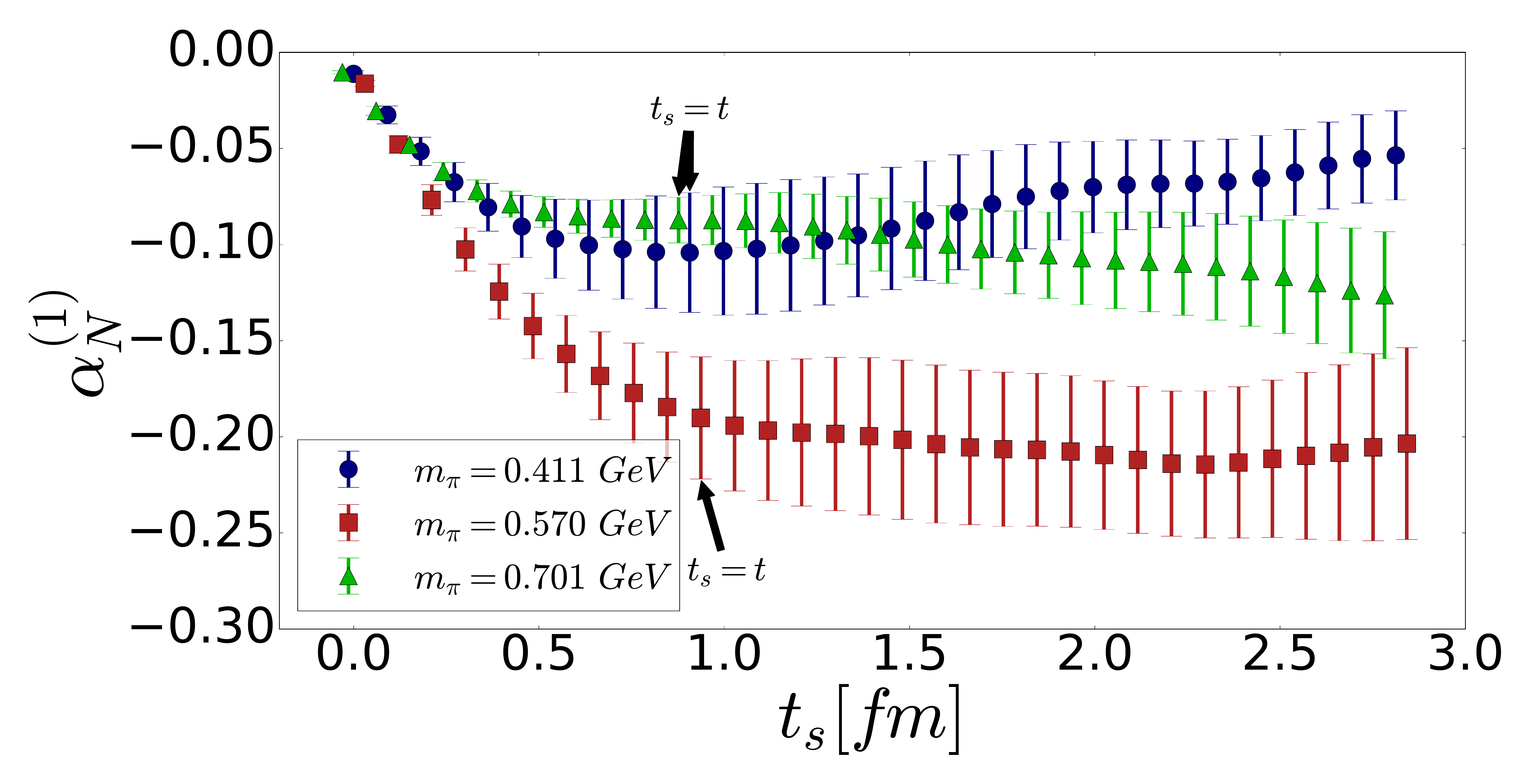}
\end{subfigure}
\quad
\begin{subfigure}{.45\textwidth}
  \centering
  \includegraphics[trim={11mm 0cm 12mm 0cm},clip,width=\linewidth]{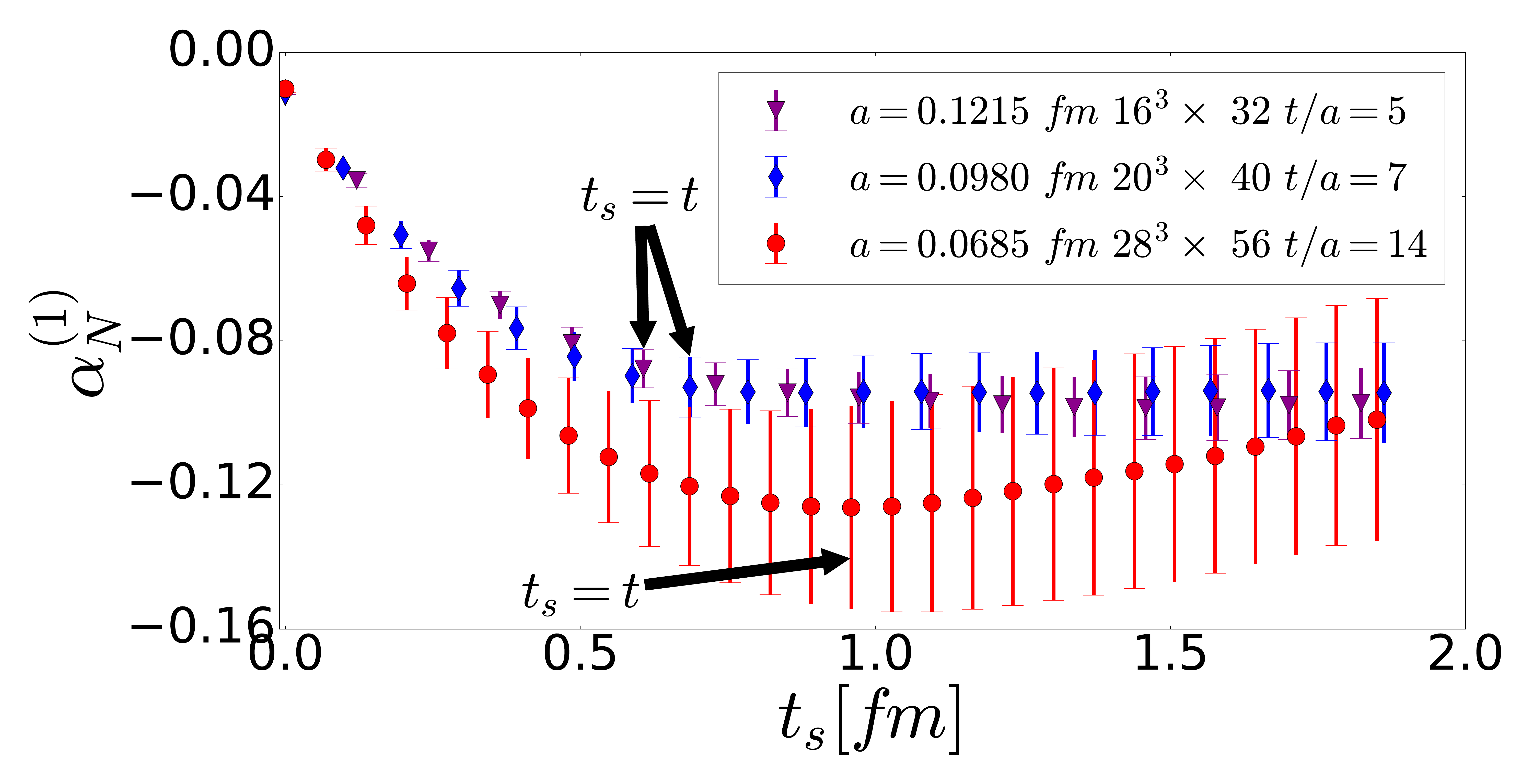}
\end{subfigure}
\caption{\label{fig:a_imp}
  \(m_{\pi}\) (left, set with $t/a=10$) and lattice spacing (right) comparisons of the improved
  nucleon mixing angle \(\alphan\) plotted against the sum parameter \(t_{s}\).
  The final point coincides with the regular nucleon mixing angle from Section.~\ref{sec:alpha}.
  The arrow indicates the location of the nucleon interpolating operator at the sink, $t$.
  }
\end{figure}

The $t_{s}$ dependence of the nucleon mixing angle $\alphan(\theta)$ is shown in fig.~\ref{fig:a_imp} for
both the $m_{\pi}$ (left) and lattice spacing (right) ensembles.
For all ensembles, we observe a saturation around the value of $t_s\simeq t$, i.e.
the values of $\alphan^{(1)}$ obtained summing up to $t_s \simeq t$ are
statistically consistent with the values obtained summing up to $t_s=T$.

\section{Improvements to the Ratio Functions\label{sec:rat_imp}}
The exact same procedure can be applied to the modified three-point correlation functions, when attempting to compute
the nucleon vector current $\mathcal{J}_{\mu}$ matrix elements. We can determine form factors with
appropriate insertions of the spin projector $\Gamma$.
We start by defining the modified three-point correlation function,
explicitly leaving in the time dependence of the
topological charge (denoted $\tau_{Q}$)
\begin{equation}
  G_{3}^{(Q)}(\vec{p}^{\, \prime},t,\vec{q},\tau,\Gamma,\tau_{Q},t_{f}) =
  \sum_{x,y}
  e^{-i  \vec{p}^{\, \prime}  \cdot \vec{x}}
  e^{i\vec{q}\cdot \vec{y}}
  \mathrm{Tr}\left\lbrace
  \Gamma
  \braket{\chi(\vec{x},t)
  \mathcal{J}_{\mu}(\vec{y},\tau)
  Q(\tau_{Q},t_{f})
  \overline{\chi}(\vec{0},0)}
  \right\rbrace\,.
\end{equation}

\begin{figure}[H]
\centering
\begin{subfigure}{.29\textwidth}
  \centering
  \includegraphics[trim={11mm 0cm 12mm 0cm},clip,width=\linewidth]{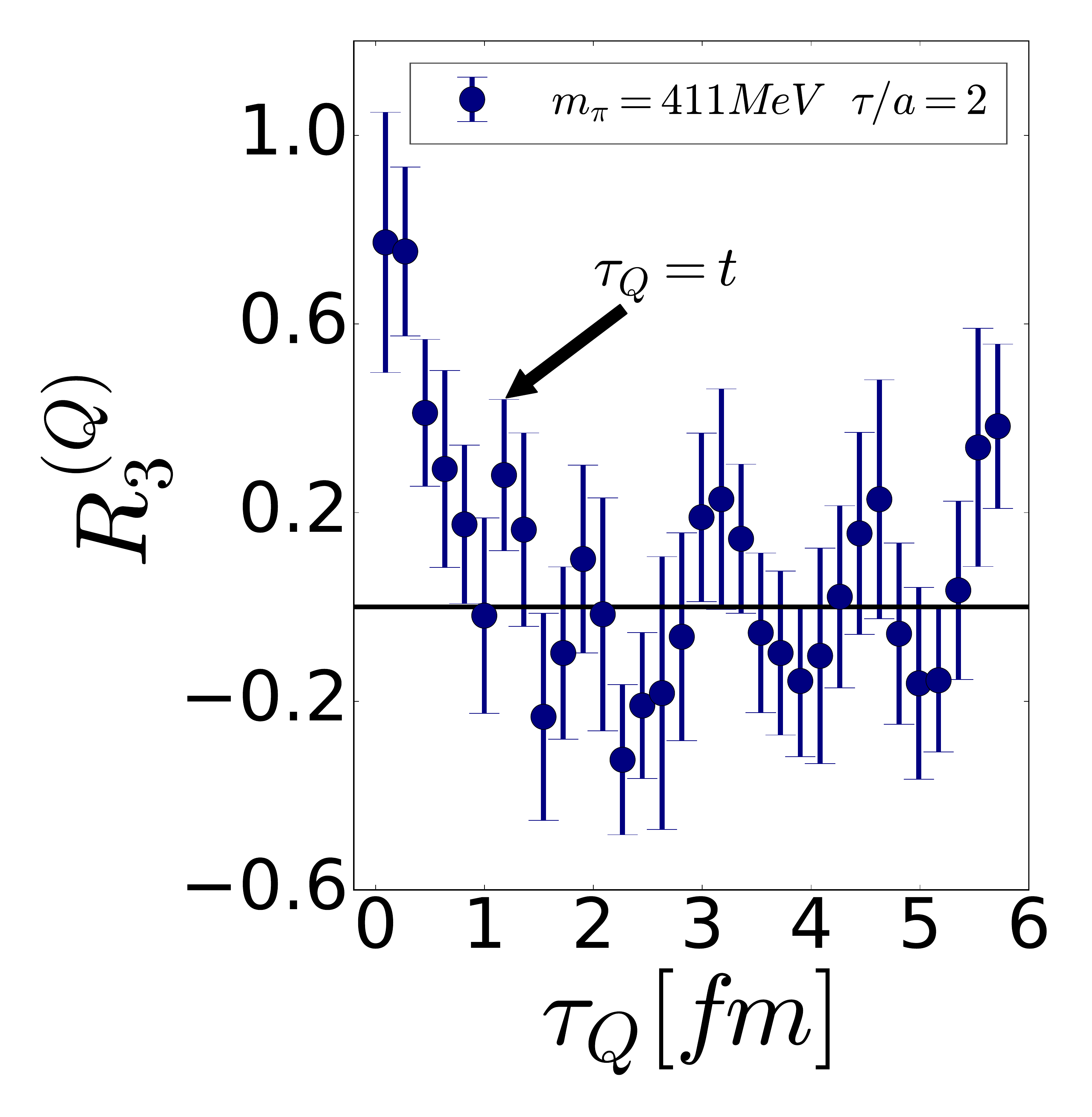}
\end{subfigure}
\quad
\begin{subfigure}{.29\textwidth}
  \centering
  \includegraphics[trim={11mm 0cm 12mm 0cm},clip,width=\linewidth]{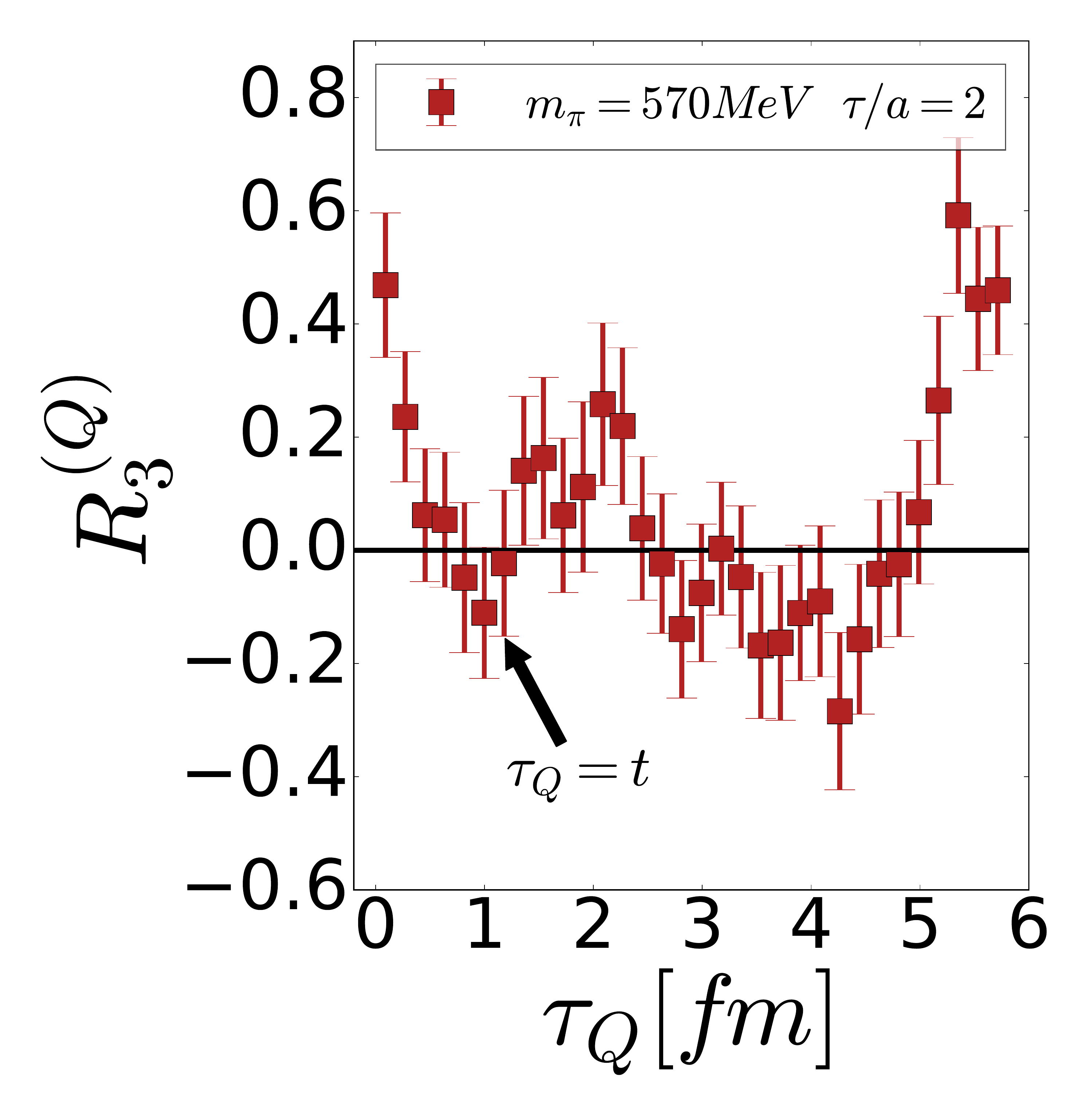}
\end{subfigure}
\quad
\begin{subfigure}{.29\textwidth}
  \centering
  \includegraphics[trim={11mm 0cm 12mm 0cm},clip,width=\linewidth]{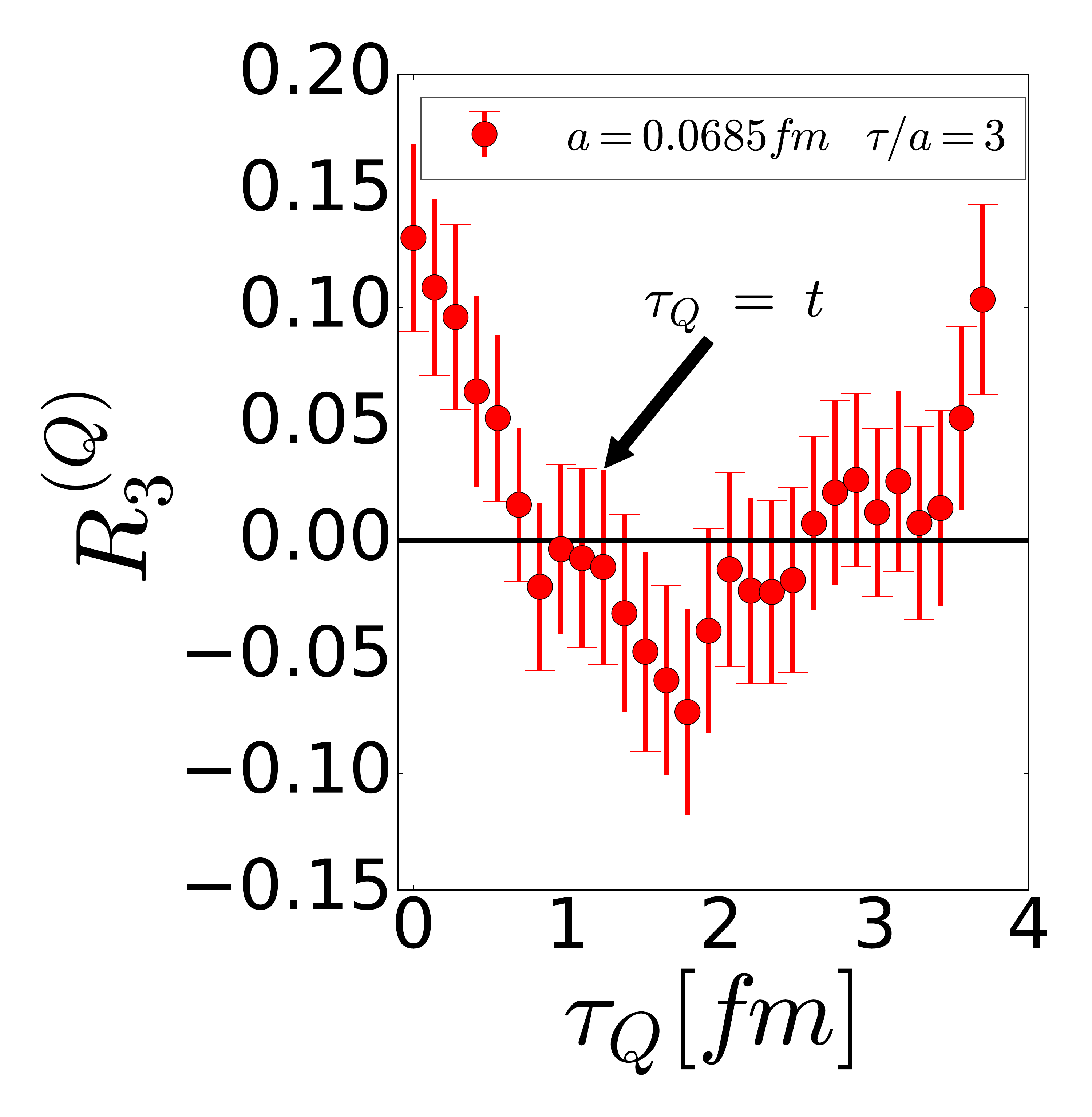}
\end{subfigure}
\caption{
  Topological charge Euclidean time dependence \(\tau_{Q}\) with respect to the ratio function \(R^{(Q)}_{3}\).
  Every even \(\tau_{Q}\) is excluded to improve visualization.
  For maximal signal, the unit of momentum \(a\vec{q}=[0,0,1]\) (blue) and \(a\vec{q}=[0,0,2]\) (red),
  the vector current with \(\mathcal{J}_{\mu = 4 }\) , projector \(\Gamma=\Gamma_{+}\gamma_{3}\gamma_{5}\), sink time \(t\)
   and current insertion time \(\tau\) (both indicated in legend) are selected.
  The left and middle and right plots are the \(m_{\pi}=\{411,570\}\)~MeV and \(a=\{0.0685\}\) fm ensembles.
  \label{fig:G3_imp}}
\end{figure}

When plotting this quantity in fig.~\ref{fig:G3_imp} we find,
for the particular nucleon matrix element in question,
that the signal occurs when the topological charge is near the source location of the nucleon (\(\tau_{Q}=0\)).
This motivates symmetrically summing $\tau_{Q}$ around $0$.

\begin{figure}[H]
\centering
\begin{subfigure}{.48\textwidth}
  \centering
  \includegraphics[trim={11mm 0cm 12mm 0cm},clip,width=\linewidth]{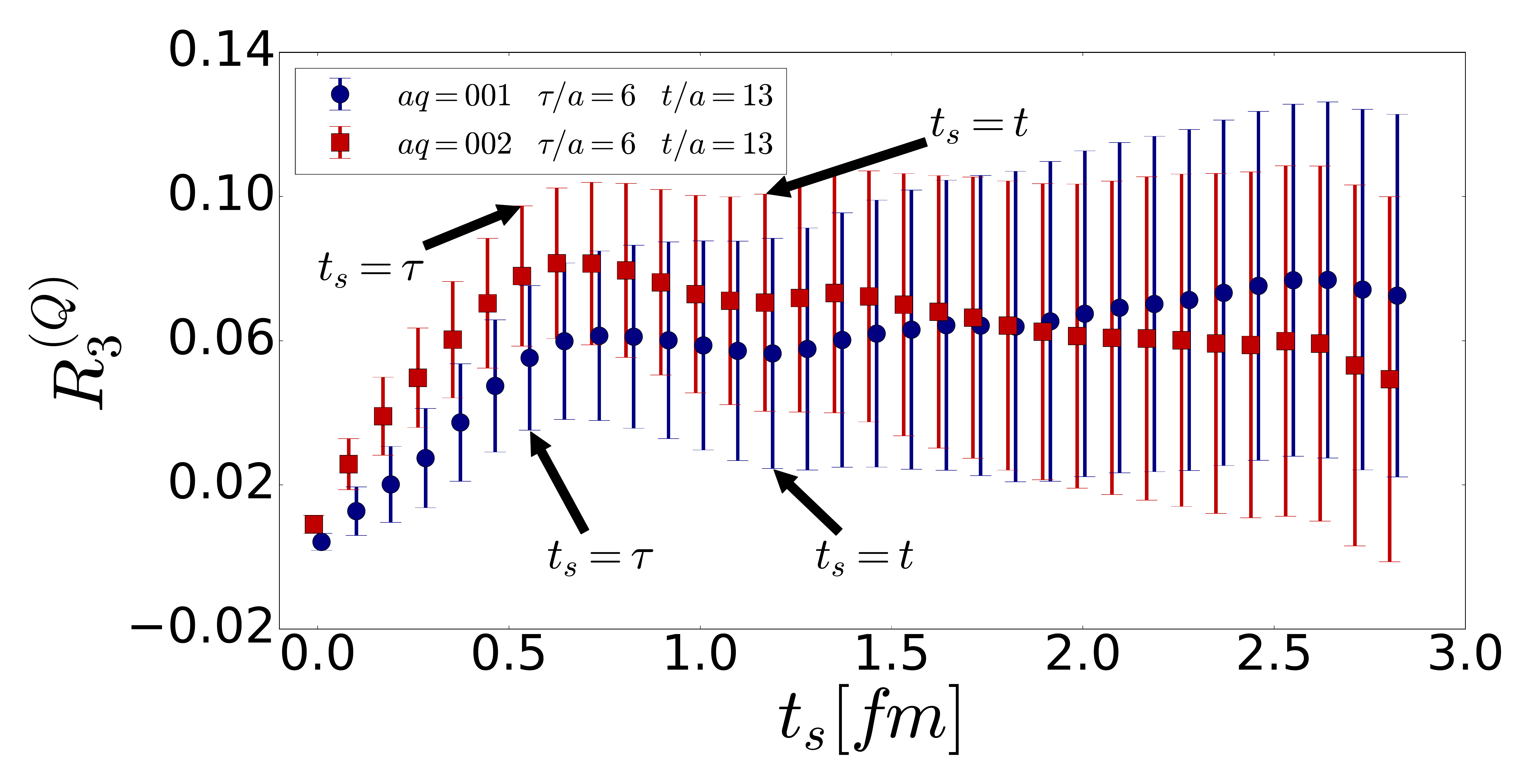}
\end{subfigure}
\quad
\begin{subfigure}{.48\textwidth}
  \centering
  \includegraphics[trim={11mm 0cm 12mm 0cm},clip,width=\linewidth]{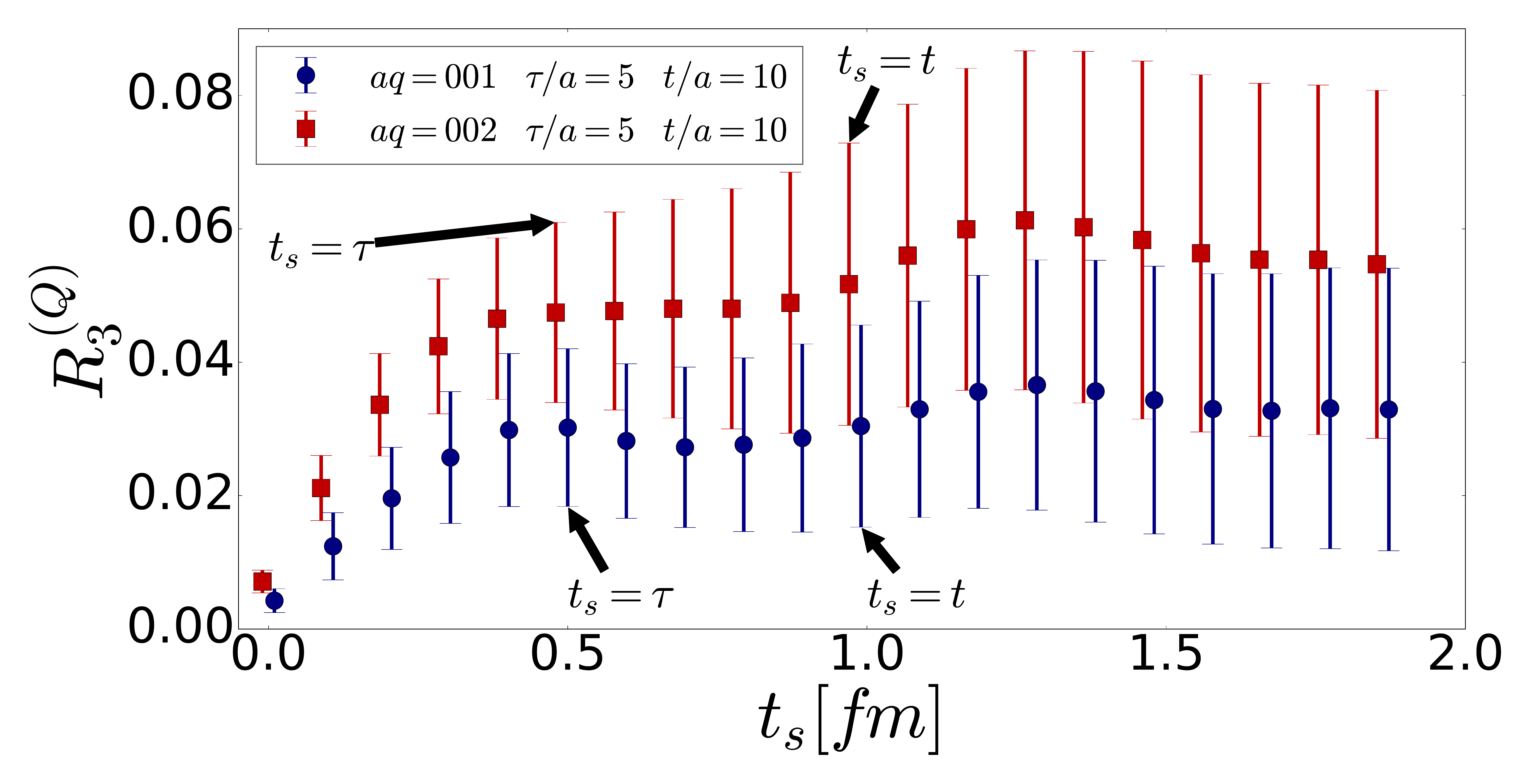}
\end{subfigure}
\caption{
    Summed topological charge Euclidean time dependence \(t_{s}\) with respect to the ratio function \(R^{(Q)}_{3}\).
  For maximal signal, the unit of momentum \(a\vec{q}=[0,0,1]\) (blue) and \(a\vec{q}=[0,0,2]\) (red),
  the vector current with \(\mathcal{J}_{\mu = 4 }\) , projector \(\Gamma=\Gamma_{+}\gamma_{3}\gamma_{5}\), sink time \(t\)
   and current insertion time \(\tau\) (both indicated in legend) are selected.
  The standard value for this quantity is obtained by taking the final \(t_{s}\) value.
  The left and right plots are the results computed on the \(m_{\pi}=570\)~MeV and \(a=0.0980\) fm ensembles.
  \label{fig:G3_imp_sum}}
\end{figure}

The quantity relevant to determine the form factors is the ``ratio function'', defined as:
\begin{eqnarray}\label{eq:RatFun}
  R^{(Q)}(\vec{p}^{\, \prime},t,\vec{q},\tau,\Gamma,\tau_{Q},t_{f}) &=
  \frac{
  G^{(Q)}_{3}(\vec{p}^{\, \prime},t,\vec{q},\tau,\Gamma,\tau_{Q},t_{f})
  }{
  G_{2}(\vec{p}^{\, \prime},t,\Gamma_{+})
  }
  \sqrt{
  \frac{
  G_{2}(\vec{p}^{\, \prime},\tau)
  G_{2}(\vec{p}^{\, \prime},t)
  G_{2}(\vec{p},t-\tau)
  }{
  G_{2}(\vec{p},\tau)
  G_{2}(\vec{p},t)
  G_{2}(\vec{p}^{\, \prime},t-\tau)
  }
  }.
\end{eqnarray}
In fig.~~\ref{fig:G3_imp_sum} we show selected results for a summed charged density
as a function of the summation range $t_s$.
The plots show that in all case the ratio functions reach a plateau for values of $t_s\simeq \tau$, i.e. the location
of the sink. Similar as was the case for $\alphan$, extending the summation up to values of $t_s=T$ it only increases the noise
of the final result.


\vspace*{-6mm}

\section{Conclusion\label{sec:con}}
The determination of nucleon two-point and three-point correlation functions are the foundation of any lattice QCD computation
of nucleon observables. When we move to a CP-violating theory with a $\theta$-term treated in a pertubative manner,
we must compute nucleon two-point and three-point correlation functions with a topological charge insertion which interacts
with the nucleon systems. As these observables suffer from severe signal-to-noise problems, improving these quantities are
of the highest priority. In this proceedings, we have explored and presented a method, based on the principle that the topological
charge will only couple to the system in question when they are ``close in Euclidean time''.


We began by studying the interaction of the topological charge with a propagating nucleon state. With the
appropriate ratio, the nucleon mixing angle can be determined from this quantity. The following section~\ref{sec:alpha_imp}
studied how the interaction between the nucleon and the topological charge is suppressed as the
topological charge is far away in Euclidean time from one of the nucleon interpolating operator. Although
the results of this study can produce an improvement on the order of $1.5$-to-$2$ times, we also observed some
statistical fluctuations in the region of exponentially suppressed signal
(where \(Q\) is "far away from the nucleon").

Lastly, the topological charge can be studied in relation to a nucleon three-point correlation function, as shown in
section.~\ref{sec:rat_imp}. After constructing the ``ratio function'', and selecting appropriate momenta and spin projectors,
the topological charge can be inserted with varying temporal location to determine where the signal is dominating.
For the parameters chosen in this paper, summing around the source location of the nucleon, produced a exponentially
convergent dependence. As the signal to noise for this observable is quite poor, we did not observe any statistically
significant disagreement between the improved and unimproved techniques. Along with this, we once again observed
even greater signal-to-noise improvements, ranging on the order of a factor $2$-to-$4$.

\bibliography{references}

%

\end{document}